REVIEW PAPER

# TEACHER AND STUDENT EXPERIENCES IN ONLINE CLASSES DURING COVID-19 PANDEMIC IN SERBIA, BOSNIA AND HERZEGOVINA AND CROATIA

[1]Amila Dautbašić , [2]Senad Bećirović

[1]*International Burch University*, Sarajevo, Bosnia and Herzegovina
[2]*International Burch University*, Sarajevo, Bosnia and Herzegovina

*Correspondence concerning this article should be addressed to Amila Dautbašić, International Burch University, Sarajevo, Bosnia and Herzegovina. E-mail: amila.dautbasic@stu.ibu.edu.ba and Senad Bećirović, International Burch University, Sarajevo, Bosnia and Herzegovina. E-mail: senad.becirovic@ibu.edu.ba*

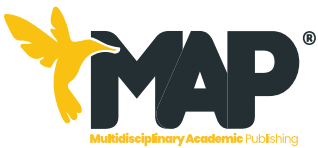

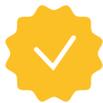

## ABSTRACT

In March 2020, the World Health Organization declared the COVID pandemic, which caused interruptions and delays in many activities, but most importantly, it led to some huge changes in education. Online teaching will prove to be the most commonly used method that should compensate for the inability to work in the classroom and allow the educational process to continue. Of course, this teaching method was not created in 2020, but it was only presented and implemented in Serbia, Bosnia and Herzegovina and Croatia with the beginning of the pandemic. In this paper, we see how these countries have faced abrupt changes in teaching, and how this change has affected students. Online teaching cannot be a mere transfer of analog content to digital; a different approach is needed in the implementation of teaching as required and offered by the digital medium, but at the same time it is necessary to preserve the basic principles of the lecturer and the curriculum. It is a call, both for teachers and students. Since this is a current and universal problem, we hope that the conclusions presented are useful.

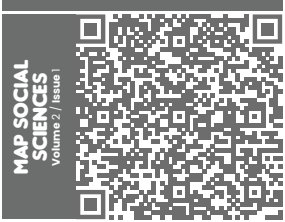

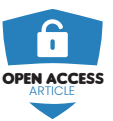

## INTRODUCTION

Online classes were mainly organized by language schools and individuals who used to teach independently until the COVID-19 pandemic started in early 2020. This type of classes was most suitable for students who were unable to physically come to live classes for a variety of reasons (Dautbašić & Saračević, 2019). Online classes were generally arranged by unofficial educational institutions since there were courses or individual classes where teachers and students applied from different countries for a certain period of time. All the technical support and infrastructure was believed to be in function, and was supposed to be well distributed to support online classes, but as it would turn out later, it was unfortunately not good enough. When it comes to online learning assets and apps for communication and social networks, it was needed to develop specific educational online tools and to use them on a much wider scale (Code et al., 2020).

Since the start of the COVID-19 pandemic in March 2020, there has been a growing number of studies revealing how teachers have encountered significant challenges in this compulsory and online education, for all (Karsenti et al., 2020), in particular regarding the adoption of this new mode of teaching which did not always correspond to their teaching practices in the classroom (Mahmood, 2020, Yaman & Bećirović, 2016). In fact, much of the research in education has focused on the challenges faced by teachers, as well as the various strategies to overcome them. Many researchers have also been devoted to the fatigue inherent in online teaching, which several refer to as "Zoom Fatigue" (Lee, 2020). This is a fatigue that many teachers experience from teaching online, often for several hours a day.

The work of Nugroho (2020) has focused more particularly on the challenges faced by language teachers. The results of their exploratory study indicate large differences related to their digital use, their digital competence and the adaptation of their teaching content to online teaching. Their results reveal, among other things, large differences in the feeling of competence of teachers, and that this is often linked to their use of digital technology in teaching in general.

Research by Kaden (2020) has focused on the workload of teachers inherent in this forced shift to distance education. Their findings clearly show that teachers, forced to teach from distance during this pandemic, have significantly increased their working hours. Finally, the work of Kim and Asbury (2020) focused more particularly on the well-being of teachers who were forced to work remotely, in the context of a health crisis. Their work details many of the challenges teachers faced, especially during the first weeks of this difficult time. The whole world faced big challenges in the time of COVID pandemic, and the aim of this research is to present the challenges that Balkan countries faced during the pandemic in terms of online learning.

## ONLINE CLASSES DURING COVID PANDEMIC IN SERBIA

E-learning and distance learning have significant advantages over traditional forms of education. Above all, they provide a great flexibility in work, because students can determine the place, time and pace of learning. This increases the efficiency of education, because it affects the work satisfaction of students and encourages their creativity. Distance learning enables a large number of students to be educated in a short time, without hiring a larger number of teachers or large classrooms and areas, so this type of education is much cheaper than the classic one. In the difficult conditions of the suspension of classical classes in schools in Serbia due to the COVID 19 virus pandemic, a large number of different programs and alternative digital ways of teaching at all educational levels were organized, the most important of which was distance learning. Although before that, distance learning in Serbia was not at a very high level, the educational system in the newly created extraordinary circumstances applied a number of measures and activities that enabled continuous and uninterrupted realization of the educational process in the new conditions (Marković, 2009).

In addition to television classes, a number of platforms were available to students, which comprehensively enabled the regular end of the school year, as well as the first online testing of an entire student generation in Serbia (Krstić et al., 2021). The organization of educational content and distance learning in Serbia in the extreme conditions of the COVID 19 pandemic confirmed the great importance of this type of education and the application of modern technologies in the implementation of teaching processes. It also opened new possibilities for their further intensive development and application in education in the future. Having in mind all the advantages, experiences and results achieved in the implementation of the distance learning process during the pandemic of the COVID 19, it is realistic to expect that this type of education will be more present in regular educational practice in the future (Marković, 2009).

The state and competent bodies and institutions contributed to that, by developing the activities started during the pandemic in the following period and creating better conditions for the development of e-education and the improvement of distance learning. And that could be achieved by providing technical preconditions, educating teachers and students for digital competencies in work and learning, as well as by creating an adequate social environment that would affirm this type of education (Dautbašić & Saračević, 2020).

Bearing in mind that the necessary transformation of educational work required a quick reaction and brought a large number of challenges, there was a fear how the leading actors in education would adapt to the changed working circumstances. The results of the research show that some teachers have successfully adapted to the different requirements that the new organization of teaching brought with it. However, a large number of teachers still face challenges in their educational work. These difficulties are reflected in the implementation of online teaching, evaluation of the teaching process and the use of digital platforms. Nearly a quarter of teachers see challenges in the organization of the educational process during the pandemic in the time load and the number of classes they teach, and over half believe that the level of administrative work that accompanies new teaching modalities is too burdensome (Stojanović, 2020).

The COVID-19 virus pandemic has undoubtedly reflected on increased fear, anxiety and the emergence of mental health problems globally. Teachers seem to have suffered a double blow from the unknown. On the one hand, like all other people, they faced a new, hitherto unknown danger, while on the other hand, they faced numerous doubts and responsibilities that their small decisions shape the long-term development outcomes of many students. Consequently, information on the challenges faced by teachers in the past is of great importance for improving the outcome of the entire educational process in a pandemic. Research findings indicate that nearly half of teachers are scared of getting infected with COVID-19 virus, but that there are no, globally, signals of excessive concern that would disrupt the psychophysical functioning of teachers. However, 10% of teachers in Serbia also experience such difficulties (such as sleep disorders, arrhythmias, etc.) (Popović et al., 2021).

The Serbian government has been very committed to ensuring the continuity of education, by switching from traditional to online teaching. They understood the current pandemic in a way as a catalyst that highlighted the need for educational change and created more flexible models. This way of thinking is an excellent basis for preparing for a potential blockage in the future.

In addition to providing a lot of tools completely free of charge, Serbia also hired various publishing houses, which enabled the free use of various textbooks and companies that provided free courses. In addition, training of lecturers for the use of digital tools is provided, and one of the most important recommendations is continuous training of lecturers for interactive teaching, with an emphasis on constant communication with pupils and students because communication is key in distance learning (Medar, 2021).

### ONLINE CLASSES DURING PANDEMIC IN CROATIA

The COVID-19 public health crisis in Croatia required a rapid change from traditional to distance learning. Digital competencies for conducting such teaching significantly affect its quality, which was evident from the results of different research where both teachers and students mostly had only basic experiences in using e-learning tools. The need for a unified model of distance learning continues to pose major challenges in the formation of any distance learning program. It is important to strive for an integrated model, while blended learning, which involves contact teaching in the presence of online content, is still considered the dominant form of online teaching, although it should only serve as a starting point for an integrated model (Picciano 2017). As Picciano concludes, the integrated model must evolve to the extent that distance education is realized because of education, and not because of the way or the distance that separates it from face-to-face teaching. Such a thing, the author predicts, will be the standard in the not-too-distant future, where every educational program should have at its core a network component, which is in itself adapted to the extent that it is invisible and does not affect the creation of educational programs. Over the last 30 years, Croatia has faced several crisis situations, which have had a major impact on all forms of teaching. Crisis situations such as the Homeland War (1990-1995), severe natural disasters (earthquakes in 2020 and 2021) and the public health crisis COVID-19, have led to some significant changes in the characteristics of teaching activities and represent valuable experience in this area (Šušak, 2021).

The COVID-19 pandemic crisis has brought to the world new problems and challenges that humanity has had and continues to face. It affected all aspects of society, including the implementation of education. By studying the relevant literature from the Croatian area, we soon realize that distance learning in Croatia is not the best developed, and that development initiatives and the offer of online curricula in Croatia are insufficiently expanded (Katavić et al. 2018).

The importance of adapting teaching materials in an online environment was presented almost 15 years ago in a paper by Bazić and Minić, where the importance of changing roles between teachers and students was also mentioned - where teachers are required to take an increasingly "pedagogical" approach and students are required an increasingly "independent" approach. Accordingly, teachers must be constantly technologically educated in order to, in addition to the classical evaluation of their students, become "organizers, developers, helpers and advisors", while at the other end of the spectrum, students should become more creative and more responsible, ie more independent (Bazić & Minić, 2007, Bećirović & Akbarov, 2016; Mašić at al., 2020).

The Ministry did not have a developed plan for the implementation of distance learning, but in parallel with the transition to distance learning in July 2020, issued a document entitled Action Plan for the implementation of distance learning for the school year 2020/2021 (Akcijski plan za sprovedbu nastave na daljinu, 2020).

The first purpose of the Action Plan for the implementation of distance learning for the school year 2020/2021. was to document the most important steps and procedures carried out in the spring of 2020 in the successful implementation of distance learning in schools and universities so that similar or improved approaches can be used in the next school and academic years and to make plans for the coming period. The Action Plan document is considered unique because it can be divided into two parts. The first part shows the stages in the implementation of the project and the manner of project implementation in the Republic of Croatia from March to September 2020, while the document was still being drafted. It is also unique because of the second part of the document, which elaborates many scenarios for future school years in order to better prepare for possible disasters around the pandemic and the best possible implementation of remote work. Croatian education, primarily teachers and students, but also all other stakeholders, has adapted relatively quickly to the new situation due to the pandemic and the inability to transfer information from school desks. The transition to the digital way of transmitting information was not a problem because Croatia started with the reform of education in 2016, and in 2017 the digital transformation was included in the reform.

Reading the above, it can be said that education in Croatia had a good basis for distance learning, which was forced due to the COVID-19 virus. COVID-19 did not initiate the digitalization of education, but only accelerated it and encouraged a major change in the educational system, which is the development and normalization of distance learning as a daily routine (Koroman, 2021).

The ability to organize distance learning during the COVID-19 pandemic has its advantages and disadvantages. The biggest disadvantage is time, because the pandemic came on abruptly and the education system was not ready to move at that speed with distance learning and its entire organization without any difficulties. As distance learning in the Republic of Croatia was not new and work was being done to improve e-learning on a daily basis, schools quickly realized various possibilities for organizing classes without taking them out of school. COVID-19 has only accelerated the "emergence" of distance learning as something that is everyday and "normal". We need to accept that technology is moving forward every day. Some countries have been using this way of teaching under normal conditions for many years and our education system is lagging behind. Online learning offers us opportunities. Opportunities are the biggest advantage of distance learning, its implementation and organization. Distance learning and its possibilities will bring new challenges to the organization of teaching, and thus more creative forms of work.

## ONLINE CLASSES DURING PANDEMIC IN BOSNIA AND HERZEGOVINA

The online teaching process in Bosnia and Herzegovina has become a novelty not only for students but also for teachers. It is only in such circumstances that most teachers began to understand that the upbringing and education of children and young people for a quality life in the 21st century must be aimed at building competencies that are necessarily different from those dominants in the past (Bećirović & Polz, 2021). The complex construct of current competencies is based on productive general and specific knowledge with which a person successfully operates and applies them outside the educational context, in everyday and professional life (Dimić, 2013). In 2005, the European Commission, as part of the Information Society and Media initiative, set out eight key competences that every European citizen should possess in order to prosper in a knowledge-based society and economy. These competencies include: the ability to communicate in native and foreign languages (Dervić & Bećirović, 2019), build core competencies in mathematics, science and technology, digital competencies, interpersonal and intercultural skills (Jarke, 2015). A significant place is occupied by information literacy skills, which became extremely important after the outbreak of the COVID virus.

It has become clear to teachers in the educational process in the schools of the Sarajevo Canton, and of course to other parts of the country, that social changes impose new obligations, roles and opportunities on the school and the teacher (Bećirović & Akbarov, 2015). It was necessary to develop a new model of school with creative-innovative foundations of paradigmatic nature, with the parallel development of pedagogical competencies of teachers (Jurčić, 2012). Therefore, a teacher in today's school needs to have developed various competencies in order to meet the requirements of today's school. Competences include knowledge, skills and personality traits, and according to Foro (2015) they are divided into key professional competencies (methodological, social and self-competence) and practical competencies and define them as combinations of knowledge, skills, attitudes, motivation and personal characteristics that enable an individual to act actively and effectively in a certain (specific) situation. According to Palekčić (2005), there are content-subject competencies, diagnostic competencies, didactic competencies, competencies in classroom management and empirical research of effectiveness that belongs to the pedagogical competencies of teachers. In the European Reference Framework, competences are defined as a combination of knowledge, practical, cognitive and social skills, and attitudes and values relevant to a particular professional and life context (European Communities, 2007).

The forced transfer of teaching to the online space in Bosnia and Hercegovina has highlighted the need for better, mostly systematic development of digital competencies of educators who in this aspect have so far been left largely to themselves. The teaching staff agrees that regular classes will not be the same as before and that computer literacy will be implied for teachers in the future. Numerous school reforms taking place in many countries around the world, which are related to the harmonization of educational systems with modern social development and building a knowledge society of the 21st century, include identifying skills needed for teachers in the information society (Bećirović & Podojak, 2017). UNESCO considers it crucial to identify information literacy indicators that should be applied in teacher education (Catts and Lau, 2008). An integral part of digital competence is digital literacy, which is defined as a basic set of skills that includes the ability to work with numbers and documents, such as word processing software and spreadsheets, web browsers, e-mail and browsers. Digital literacy refers to the ability to read and understand hypertext or multimedia texts, and includes comprehension of images, sounds, and text (Petzanet, 2019).

Not going so far into the future, this time marked by the COVID epidemic only gave an indication of what the future holds and what will be necessary to adopt a teacher to keep up with the times and continue to meet important didactic and methodological aspects of the teaching process. The era of digital media is putting new pressure on schools. Schools should prepare students and young people for qualitatively changed opportunities in society and their future roles as workers and citizens. The change in the world of work should be accompanied by a change in the content, form and structure of learning. An important feature of this fresh learning culture is that learning is changing almost on a daily basis and that it needs to ensure its own constant variability. In this sense, a 21st century teacher, in order to be a literate teacher of the digital age and competent to lead the teaching process, will need to have visual and information skills and multicultural literacy and global awareness. Therefore, the transformation of the learning and teaching process is necessary in order to improve them and harmonize them with the current time, and even with the projected future needs (Urusa & Malik, 2013).

In primary and secondary schools in Sarajevo Canton, classes during the COVID epidemic took place through online platforms. This was preceded by the suspension of the regular teaching process in the classrooms, which is the product of an order issued in accordance with the decisions of the Government of the Federation of BiH. It is a matter of declaring a state of natural or other disaster on the territory of Bosnia and Herzegovina, and due to the danger of spreading an infectious disease caused by COVID, as well as in accordance with orders of the Federal Civil Protection Headquarters and the KS Government Decision on declaring a state of natural and other disaster. Soon, in order to implement online teaching, professional teams were formed, and by orders of the relevant ministry, recommendations were issued to all schools on new work circumstances. Thus, online classes were treated as regular classes that the students successfully completed on June 2020, and in which there was no compensation for classes (Sekulić et al., 2021).

By definition, online classes are an educational process in which teacher and student are separated and in which educational content is delivered to remote locations using information and communication technologies that include written correspondence, text and hypertext, graphics, audio and video media, direct instructions via telecommunication channels (so-called online learning), application of audio and video conferencing systems, interactive TV. Many teachers have realized that the use of the latest information and communication achievements during the educational process breaks the tradition in which teaching is a one-way process where the teacher transfers knowledge, which students passively adopt and then reproduce as learned (Derviš & Bećirović, 2020; Glazier, 2016).

However, the use of information and communication technologies has helped teachers to create a stimulating learning environment within their subject in a way that makes teaching more creative, challenging, interesting, functional and purposeful. Most primary and secondary schools in Sarajevo Canton used the Office 365 Microsoft Teams platform, which proved to be very practical in the online teaching process. On the same platform, it is possible to form so-called "classrooms" that can have different purposes (Bećirović et al., 2019). Thus, one classroom can consist of one class with a teacher of a specific subject. Also, for the purpose of fulfilling all pedagogical and psychological obligations, classrooms can be composed exclusively of teaching and pedagogical staff, i.e. Department or Teachers' Councils. It enables visual and textual exchange of contents in order to fully realize the teaching process. It is important that the same platform enables the archive of all documents, data and communication, which can always be an excellent proof of the manner, quality and realization of the online teaching process (Martin & Tapp, 2019).

Teachers have different experiences with the use of this platform and the online teaching process in general. However, everyone agrees that this is an important step towards future modalities of the teaching process, and that for many teachers this was the first serious experience with information and communication. technology in teaching (Dautbašić & Saračević, 2020).

During the analysis of online teaching, it proved imperative that it is absolutely necessary to adapt the teaching faculties to the requirements of the modern teaching process, which, in addition to direct school teaching, must also include distance learning. Therefore, it is necessary to study subjects more frequently in the study of teacher training, which will satisfactorily treat the application of IT tools and technological achievements (Song et al., 2016). This will at all times, especially the information age, put the teacher in the place he deserves, and that is to be able to educate students in the essence of the object and subject of study of their field, either through online teaching or other forms of teaching process. We should not forget that the best teachers are those whose students know best (Spahić, 2015). Through a concrete analysis of online teaching in Sarajevo Canton and educational and demographic opportunities in this canton, it is possible to identify several important things. Sarajevo Canton, due to the earlier introduction of the diary, has a clearer picture in the direction and development of online teaching process than other cantons, has the most IT-literate population, records the lowest trends in reducing the number of students in primary and secondary schools and reducing the number of first-graders. natural increase as a very important factor in the development of education (Sekulić et al., 2021).

Regardless of what is considered the most favorable canton in these parameters, the general picture of the situation indicates lagging behind the average of developed countries in Europe, especially when it comes to general computer literacy, the highest graduation rate and readiness to accept challenges with online teaching (Vasiljević, 2020).

## CONCLUSION

Understanding the student's point of view is the key to creating an effective school environment in its student-centered meaning aimed at performing learning processes at any level (Gover et al., 2019). It should not be forgotten, in fact, also from a perspective and future didactic transformation, that the school is, and should always remain, a social environment called to fulfill the primary role of training students in the pedagogical meaning of the term it holds, therefore, education is taken into account, but also the pivotal aspect of education for life. There are foundations on which a future school system must be built capable of guaranteeing an effective combination of traditional elements, devoted to the care of "student-student" and "student-teacher" relationships, and of those technologies aimed at enhancing students' digital skills and optimizing intangible resources, as well as the spatial and temporal conditions that these tools make possible to obtain (Dautbašić & Saračević, 2020). An optimal model in this direction needs time to be structured first, and then implemented, and everyone is called upon not to miss the opportunity for this unexpected and forced didactic transformation to become innovation: families, political institutions, principals, teachers, and students - they all have the moral obligation to commit themselves so that the innovations introduced are perfected and become innovative resources at the service of tradition. Unfortunately, as we still live in a time of pandemic, we are faced with various risks and uncertainties. Thus, there are no best practices or a single model for colleges and universities to follow. The education systems of countries around the world, including Serbia, Bosnia and Herzegovina and Croatia, faced various problems even before the pandemic, so they should see the COVID-19 pandemic as a chance to make radical changes to their education systems (Dervić & Bećirović, 2020).

The recommendation for future studies is to focus on encouraging the general public to establish successful strategies for dealing with current and future challenges in education, as well as raising awareness that changes in the education system are necessary, not because of COVID-19, but because of the inevitable changes the world is about to face in terms of digitalization and presumably the evolution of online education. Digitized learning should become a normal practice in regular circumstances. All educational institutions must be ready for the difficult path that will follow after the end of the pandemic, because their decisions will shape and direct the future of their students.